\documentclass[journal]{IEEEtran}

\usepackage{amsmath,amsfonts,amssymb}
\usepackage{array}
\usepackage[caption=false,font=normalsize,labelfont=sf,textfont=sf]{subfig}
\usepackage{textcomp}
\usepackage{stfloats}
\usepackage{url}
\usepackage{verbatim}
\usepackage{graphicx}
\usepackage{cite}
\usepackage{booktabs}
\usepackage{multirow}
\usepackage{microtype}
\usepackage{mathtools}
\usepackage{adjustbox}
\usepackage{multicol}
\usepackage[ruled]{algorithm2e}
\usepackage{siunitx, etoolbox}
\usepackage{url}
\usepackage{overpic}

\newcommand{\T}{\operatorname{T}}       
\newcommand{\Her}{\operatorname{H}}     

\begin{document}
	
\title{Detecting Malicious Pilot Contamination in Multiuser Massive MIMO Using Decision Trees}

\author{Pedro Ivo da Cruz,
	Dimitri Silva,
	Tito Spadini,
	Ricardo Suyama,
	and Murilo Bellezoni Loiola
	\thanks{The authors are with the Engineering, Modeling and Applied Social Sciences Center, Federal University of ABC, Santo André, SP, Brazil.}%
	\thanks{This work was financed in part by the Coordenação de Aperfeiçoamento de Pessoal de Nível Superior - Brasil (CAPES) - Finance Code 001 -, the National Council for Scientific and Technological Development (CNPq) -, and the São Paulo Research Foundation (FAPESP) under grant \#2020/09838-0 (BI0S - Brazilian Institute of Data Science).}
}

\maketitle

\begin{abstract}
	Massive multiple-input multiple-output (MMIMO) is essential to modern wireless communication systems, like 5G and 6G, but it is vulnerable to active eavesdropping attacks.
	One type of such attack is the pilot contamination attack (PCA), where a malicious user copies pilot signals from an authentic user during uplink, intentionally interfering with the base station's (BS) channel estimation accuracy.
	In this work, we propose to use a Decision Tree (DT) algorithm for PCA detection at the BS in a multi-user system.
	We present a methodology to generate training data for the DT classifier and select the best DT according to their depth.
	Then, we simulate different scenarios that could be encountered in practice and compare the DT to a classical technique based on likelihood ratio testing (LRT) submitted to the same scenarios.
	The results revealed that a DT with only one level of depth is sufficient to outperform the LRT.
	The DT shows a good performance regarding the probability of detection in noisy scenarios and when the malicious user transmits with low power, in which case the LRT fails to detect the PCA.
	We also show that the reason for the good performance of the DT is its ability to compute a threshold that separates PCA data from non-PCA data better than the LRT's threshold.
	Moreover, the DT does not necessitate prior knowledge of noise power or assumptions regarding the signal power of malicious users, prerequisites typically essential for LRT and other hypothesis testing methodologies.
\end{abstract}


\section{Introduction}

Massive multiple-input multiple-output (MMIMO) is essential to modern wireless communication systems, like 5G and 6G.
It uses large arrays of antenna elements at the transmitter and receiver to simultaneously serve multiple users, resulting in higher spectral efficiency and increased system capacity compared to devices with fewer antenna elements.
MMIMO systems also provide more extensive coverage and improved robustness to fading~\cite{Jain2022}.

\begin{figure*}[ht]
	\centering
	\includegraphics[width=\linewidth]{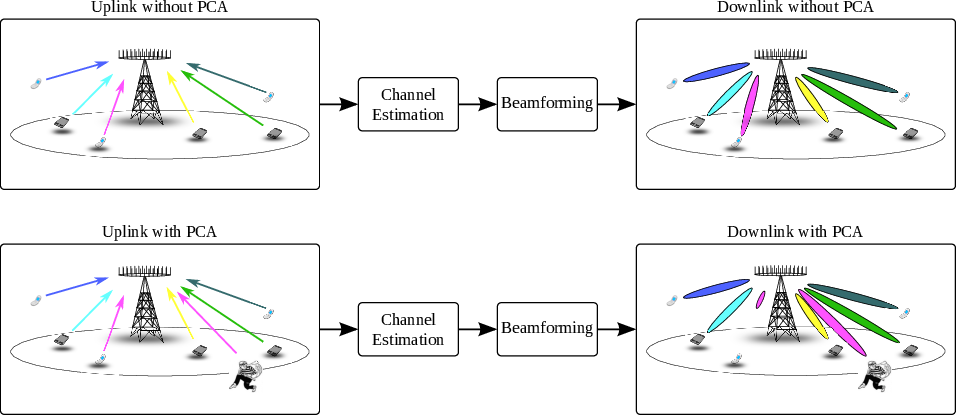}
	\caption{In a scenario without pilot contamination, the users send pilot signals to the BS, which uses this information to estimate the channels and perform beamforming. With PCA, the eavesdropper sends the same pilot signal to the BS, who will estimate the eavesdropper's channel as legitimate so that information will leak during DL.}
	\label{fig:MMIMO}
\end{figure*}

MMIMO systems achieve massive connectivity, supporting a large number of users in a cell, through the use of beamforming techniques~\cite{Abohamra2023, Karim2023, Hu2022}.
These techniques rely on Channel State Information (CSI) between users and the base station (BS) to create multiple beams, which can be directed toward different users.
To obtain the CSI, users broadcast pilot signals in an uplink (UL) transmission.
The BS then estimates the CSI for each user and uses beamforming to direct the signals to multiple users simultaneously, with each beam carrying information intended for a specific user.
This process is illustrated at the top of Fig.~\ref{fig:MMIMO}.

On the other hand, wireless communication systems are inherently susceptible to passive eavesdropping~\cite{Hamamreh2019}.
To protect against these attacks, physical layer security (PLS) techniques have been developed to safeguard wireless networks against passive eavesdropping attacks, where an eavesdropper only monitors transmitted signals to obtain confidential information~\cite{Hamamreh2019}.
Using beamforming in MMIMO can prevent passive eavesdropping attacks, making them more challenging to execute~\cite{Ramezanpour2023}.

However, active eavesdropping poses a challenging problem, as eavesdroppers can transmit signals to interfere with legitimate communication and steal confidential information~\cite{chorti2022}.
In MMIMO systems, one of the primary attacks of this type is the pilot contamination attack (PCA), when an eavesdropper transmits false pilot signals during UL, intentionally interfering with the BS's ability to estimate the CSI~\cite{Irram2022} accurately.
Since pilot signals are typically publicly known, the PCA consists of the eavesdroppers impersonating a legitimate user, which is accomplished using the same pilot signal as the legitimate user.
Significant information leakage may occur depending on the eavesdropper's transmission power, since the beam intended for the user under attack mostly shifts to the eavesdropper, as shown in the bottom of Fig.~\ref{fig:MMIMO}.

Hence, it is important to develop mechanisms to detect the presence of PCA in the UL signal, so that the BS can take countermeasures.
In the following section, we review some of the PCA detection techniques in the literature.

\subsection{Related Work}

Pilot contamination is a common problem in cellular systems due to the reuse of pilot signals in adjacent cells.
As users usually broadcast these signals, some residual pilot signals from these users in adjacent cells can reach the BS, which can interfere with the BS’s ability to estimate the CSI accurately.
This leads to performance degradation, reduced spectral efficiency, denial of service for some users, or even information theft.
However, reliable techniques can deal with these interfering signals~\cite{sedghi2022, dey2022, akhlaghpasand2017}. 

In PCA, however, the eavesdropper usually transmits the signal with high power. Therefore, techniques capable of dealing with contamination from adjacent cells might not work properly, preventing the BS from properly estimating the CSI and transmitting information to legitimate users.
Hence, it is important to develop techniques to detect the presence of PCA.

PCA detection techniques can be classified into two categories: 1) detection at the BS; and 2) cooperative detection.

Several works in the first category follow a hypothesis testing approach~\cite{kapetanovic2015, Gao2019, Xu2020, Akhlaghpasand2018, Hassan2020, Xu2021}, exploring the information provided by the received signal autocovariance matrix to detect PCA.
In~\cite{kapetanovic2015}, \cite{Gao2019} and~\cite{Xu2020}, different test statistics are proposed, but they are only capable of dealing with a simplistic scenario, with one user connected to the BS and a single eavesdropper attempting PCA.

Additional assumptions are made to develop the detection method for more realistic scenarios.
For example, \cite{Akhlaghpasand2018} considers a multi-user scenario with an eavesdropper equipped with multiple antennas and assumes that the number of available pilot signals is greater than the number of users in the cell.
In this case, the BS projects its received signal onto the unused pilot signals, assuming that the resulting signal only contains noise without PCA and a correlated signal with PCA.
A general likelihood ratio test (LRT) is subsequently developed for detection.
Other examples are presented in~\cite{Hassan2020} and~\cite{Xu2021}, which develop PCA detection algorithms based on LRTs.
In the former, the instantaneous energy of the channel estimate is used as a test statistic, but makes an unrealistic assumption that the eavesdropper and legitimate users always transmit with the same (known) power; in the latter, PCA detection relies on the manipulation of the pilot signals.

Following a different approach, the system proposed in~\cite{Banaeizadeh2021} uses a generative adversarial network (GAN) structure to identify the presence of PCA, which was extended in~\cite{Yadav2024}.
During the training process, the system learns the features of legitimate signals.
As a result, after training, it can detect the presence of a PCA by identifying changes in the originally learned pattern.
The work in~\cite{wang2018} uses a support vector machine (SVM) to detect PCA in an MMIMO system with a non-orthogonal multiple access scheme.
However, this work does not provide a method to obtain a threshold to determine the presence of PCA, and the analysis is made by varying a threshold and showing the false alarm by detection probability curves.

In \cite{Wang2020}, the authors propose a machine learning (ML)-based algorithm for PCA detection in Non-Orthogonal Multiple Access (NOMA) systems, employing a logistic regression classifier. In \cite{Wang2021}, PCA is analyzed in the context of Grant-Free IoT Networks, an emerging technology for massive connectivity with low latency in 5G massive machine-type communications (mMTC).
In this latter case, a method based on convolutional neural networks (CNNs) is proposed for PCA detection.
In fact, ML approaches have emerged as a significant trend in addressing PLS challenges, also offering promising solutions for other related problems, such as adversarial attacks on power allocation methods for MMIMO systems~\cite{Sahay2023}, pilot spoofing attack (PSA)~\cite{Wang2024}, jamming attack in the beam training procedure in 5G mmWave Networks \cite{DinhVan2023}, among others~\cite{sharma2023, zhang2024, Guo2024}.

On the other hand, cooperative PCA detection involves legitimate users in the PCA detection procedure.
For example, in~\cite{Kapetanovic2014, Zeng2021}, additional information is transmitted between the BS and legitimate nodes during the downlink (DL) stage, and the nodes perform PCA detection.
In~\cite{Kapetanovic2014} the BS applies beamforming to transmit a value previously known by the legitimate user.
When PCA is present, the legitimate user receives less than the agreed-upon value and can detect contamination.
In~\cite{Zeng2021}, the proposed detection strategy involves the BS sending two specific signals, constructed using CSI, to legitimate users.
The legitimate user uses this information to design the decision metric and determine the presence of PCA.
Nonetheless, in both cases, the eavesdropper can also contaminate the DL transmission, thus preventing accurate PCA detection by the user.

A different cooperative approach was proposed in~\cite{wang2023}.
Unlike the previous cooperative methods, PCA detection is handled by the BS.
However, the UL is assisted by a relay node that transmits its pilots, which enables channel estimation between the relay and BS.
This new information is used to create the PCA detection algorithm.
However, similarly to~\cite{Hassan2020}, this method requires knowledge of the variances of the noises in the signals and assumes the eavesdropper transmits with unitary power.

\subsection{Contributions}

The problem of detecting PCA by an active eavesdropper can be interpreted as a classification problem that can be solved using Machine Learning (ML) algorithms.
ML has been recently investigated to solve problems in wireless communication systems~\cite{Singh2022, Liu2022, Restuccia2020, Bjornson2020}.
In MMIMO systems, ML techniques have been proposed for channel estimation~\cite{Sadeghi2023, Fang2023, Guo2023, Le2021, Demir2020, He2018}, for precoding and beamforming~\cite{Hu2022, Lin2020, Girnyk2021}, for energy efficiency management~\cite{Hoffmann2021}, power allocation~\cite{dossantos2022} and also for PLS~\cite{sharma2023}.

This work proposes using ML to perform PCA detection at the BS in a multi-user MMIMO system.
More specifically, we use a Decision Tree (DT) classifier to detect PCA based on the instantaneous energy of the estimated channel.
We focus on detecting PCA at the BS to reduce the transmission overhead in the network and avoid additional processing at the users, which often have computational and energy constraints.
The contributions and main results of this work are summarized as follows:
\begin{itemize}
	\item We develop a DT classifier for PCA detection at the BS, requiring no prior knowledge  that might be unknown to the BS, such as the noise power or the eavesdropper transmission power;
	\item We present a methodology to generate the data and train the DT for PCA detection;
	\item The DT is compared to an LRT method, which requires the knowledge of the noise power and the eavesdropper transmission power to compute the test statistic;
	\item The resulting DT model results in a high probability of detection in low signal-to-noise ratio (SNR) regimes and small values of eavesdropper transmit power;
\end{itemize}

The remainder of the paper is organized as follows: Section~\ref{sec:mimo_model} presents a mathematical overview of MMIMO modeling, covering channel estimation and eavesdropper characteristics.
In Section~\ref{sec:model}, we detail the proposed detection model, including data generation, model training, performance metrics, and benchmark algorithms used for comparison.
In Section~\ref{sec:results}, we present the results obtained, examining the impact of important factors such as the SNR, number of connected users, eavesdropper power, and number of antennas at the BS.
Finally, in Section~\ref{sec:conclusion}, we conclude and outline potential avenues for future research.

\section{MMIMO Signal Model and Problem Formulation}
\label{sec:mimo_model}

This work considers a system comprising an MMIMO BS equipped with a linear uniform array featuring $ M $ antennas and $ K $ single-antenna users.
The block fading channel model is considered, wherein the channel impulse response remains constant throughout a transmission block containing the UL followed by the DL signal.

The $ k $-th user sends $N$ pilot symbols to the BS, denoted as a row vector $ \mathbf{x}_{k} = \left[ x_{k}(0), \ x_{k}(1), \ \cdots, \ x_{k}(N-1) \right] $, where $ x_{k}(n) $ represents the pilot symbol transmitted at the discrete time $ n $.
Assuming that the $ k $-th user transmits with power $ P_{k} $, the discrete-time and base-band signal received at the $ m $-th antenna of the BS in the UL can be expressed as
\begin{equation}
	\mathbf{y}_{m} = \sum_{k=0}^{K-1} \sqrt{P_{k}}\ h_{km}\ \mathbf{x}_{k} + \mathbf{v}_{m},
\end{equation}
where $ h_{km} $ is the complex channel gain between the $ k $-th user and the $ m $-th antenna, and $ \mathbf{v}_{m} $ is the additive white Gaussian Noise (AWGN) row vector with $\mathbf{v}_{m} \sim \mathcal{CN}(0, \sigma^{2} \mathbf{I}_{N}) $.
Here, the envelope of $ h_{km} $ follows a Rayleigh distribution, i.e., $ h_{km} \sim \mathcal{CN}(0, \sigma_{h}^{2}) $.

The signal received by each antenna can be organized in a matrix $\mathbf{Y} = \left[ \mathbf{y}_{0}^{\T}, \ \mathbf{y}_{1}^{\T}, \ \cdots, \ \mathbf{y}_{M-1} ^{\T}\right]^{\T} $, where $ (\cdot)^{\T} $ denotes transposition. Therefore, by defining the channel vector as $ \mathbf{h}_{k} = \left[ h_{1k}, \ h_{2k}, \ \cdots, \ h_{(M-1)k} \right]^{\T} $, the received signal can be written as
\begin{equation}
	\label{eq:bs-received-signal}
	\mathbf{Y} = \sum_{k=0}^{K-1} \sqrt{P_{k}}\ \mathbf{h}_{k}\ \mathbf{x}_{k} + \mathbf{V},
\end{equation}
with $ \mathbf{V} = \left[ \mathbf{v}_{0}^{\T}, \ \mathbf{v}_{1}^{\T}, \ \cdots, \ \mathbf{v}_{M-1}^{\T} \right]^{\T} $.

It is important to highlight that this model is representative, although simple.
It has been widely used in works regarding PCA and PSA~\cite{kapetanovic2013, kapetanovic2015, Hassan2020, Wang2020, Zeng2021, Xu2020, wang2023} and is the model used in the MMIMO seminal work~\cite{marzetta2010}.

\subsection{Pilot Contamination Attack Model}

The eavesdropper may intentionally degrade the channel estimation and potentially change the direction of the beam of an authentic user by sending pilot symbols to the BS.
To accomplish that, the eavesdropper sends the pilot symbols $\mathbf{x}_{e} $ with power $P_{e}$, thus~\eqref{eq:bs-received-signal} becomes
\begin{equation}
	\label{eq:bs-received-signal-pca}
	\mathbf{Y} = \sum_{k=0}^{K-1} \sqrt{P_{k}}\ \mathbf{h}_{k}\ \mathbf{x}_{k} + \sqrt{P_{e}}\ \mathbf{h}_{e}\ \mathbf{x}_{e} + \mathbf{V},
\end{equation}
where $ \mathbf{h}_{e} = \left[ h_{1e}, \ h_{2e}, \ \cdots, \ h_{(M-1)e} \right]^{\T} $ is the channel vector containing the channel gains between the eavesdropper and the BS, whose envelope also follows a Rayleigh distribution, just such as $ h_{km} $.

\subsection{Channel Estimation}

The channel taps between the $ k $-th user and the BS can be estimated by the BS using least squares (LS) estimation, given by
\begin{equation}
	\label{eq:ls-estimation}
	\hat{\mathbf{h}}_{k} = \mathbf{Y} \frac{\mathbf{x}_{k}^{\Her}}{\sqrt{P_{k}}\ \lVert \mathbf{x}_{k} \rVert^{2}},
\end{equation}
provided that the pilot signals are orthogonal to each other, i.e., $ \mathbf{x}_{k} \mathbf{x}_{l}^{\Her} = 0 $ for $ k \neq l $.

It is important to highlight here that the accuracy of the channel estimation is directly dependent on the power of the noise and the power of the eavesdropper pilot signal.
Suppose the eavesdropper is attempting to compromise the channel of the $k$-th user. Then, the channel estimate is decomposed as follows
\begin{equation}
	\hat{\mathbf{h}}_{k} = \mathbf{h}_{k} + \sqrt{\frac{P_{e}}{P_{k}}} \ \mathbf{h}_{e} + \mathbf{V}\frac{\mathbf{x}_{k}^{\Her}}{\sqrt{P_{k}}\ \lVert \mathbf{x}_{k} \rVert^{2}},
\end{equation}
which shows that a larger $P_{e}$ leads to a larger channel estimation error, and an increased likelihood of the BS considering the eavesdropper channel as legitimate.

\section{Detection Model and Decision Tree}
\label{sec:model}

The PCA detection problem can be viewed as a classification problem for ML techniques.
In a classification problem, each observation or sample can be described by a set of specific characteristics or data attributes, also called ``features'', represented mathematically by a vector structure.
Each position in the vector of attributes corresponds to a dimension in a hyperspace of data, enabling each sample to be represented as a point in that hyperspace.

Classifiers are ML algorithms that receive the set of attributes of a new sample and use defined criteria to identify the most compatible class based on its position in the data hyperspace.
A classifier can be described as a function that outputs a class name, or a number that represents the class, based on a vector of input attributes.
Thus, the PCA detection problem can be defined as follows: using a classifier trained with labeled data, the instantaneous energy of the channel estimate, $E = \lVert \hat{\mathbf{h}}_{k} \rVert^{2}$, and the number of users, $K$, are used to classify a sample as either PCA or non-PCA.
The classifier model identifies which class the sample is most similar to based on its vectorial attributes.

This work proposes a Decision Tree (DT) based method for PCA detection.
The DT is a well-established, non-parametric classifier that is relatively simple yet sufficiently efficient in numerous scenarios.
DTs extract rules from the attributes of training samples to partition input data into cases of a single class.
The DT receives a set of attributes corresponding to a class as input.
It examines each attribute to determine which class the attributes belong to.
The first node analyzes one attribute and, depending on its values and a set of rules, follows a path to a successor node that evaluates another attribute.
This process repeats until a final node is reached, determining the class of the set of attributes.

The choice for the DT is straightforward.
The set of rules in a DT consists mostly of threshold comparisons.
This simplifies the detection procedure once the DT is trained and the thresholds are defined, reducing the computational complexity of the PCA detector during deployment and yielding fast decisions.
This work will show that a DT of depth one, i.e., a DT that uses a single threshold, can efficiently detect PCA.
This helps achieve faster transmissions and the 5G requirements of high throughput and reliability.

\begin{figure}[ht]
	\centering
	\includegraphics{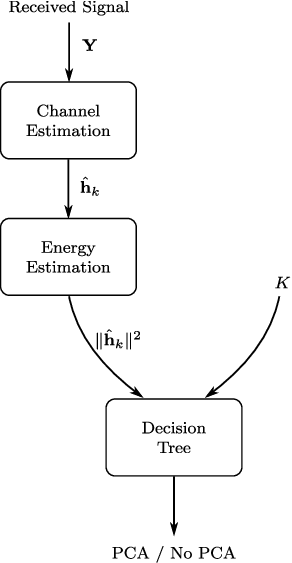}
	\caption{PCA detection mechanism.}
	\label{fig:dt-scheme}
\end{figure}

The DT must undergo training using labeled data to determine the rules and attributes evaluated in each node and the number of nodes.
These data should include the features to be used as input: the number of connected users ($K$) and the instantaneous energy of the channel estimate ($E$).
The channel estimate can be obtained using the simple LS estimator~\eqref{eq:ls-estimation}.
Fig.~\ref{fig:dt-scheme} illustrates the overall process for detecting PCA.

The dataset should include various samples for distinct operating conditions such as different SNR values, different eavesdropper transmission power ($ P_{e} $), and different numbers of users ($ K $), among others.
In the following section, we discuss the generation of this dataset.

\subsection{Data for Training and Evaluation}
\label{sec:dataset-methodology}

As with all supervised ML techniques, a dataset is necessary to train the classifier.
It is crucial to highlight that the data used in the training stage must represent the features to be encountered when the system is deployed.
Therefore, constructing the training dataset requires careful attention.
In this paper, we create the dataset for training and testing our system by simulating the UL signal transmissions.

In summary, we generate the training and evaluation datasets for a given number of BS antennas, $M$.
This is because data for different numbers of antennas is unnecessary since that quantity remains constant after the deployment of that specific BS structure.
To generate the data, we set the users' transmission power to unity and the number of transmitted pilot symbols during the UL stage to $ N = 300 $.
Next, we simulate the UL transmission for various SNR, $P_{e}$, and $K$ values, according to the model presented in Section~\ref{sec:mimo_model}.
We consider flat-fading channels whose coefficients are randomly generated for each sample in the dataset.
The envelope of these coefficients follows a Rayleigh distribution, with the coefficients independently and identically distributed.
Each sample on the dataset contains the features extracted from the simulation of the received signal in one transmission block given by~\eqref{eq:bs-received-signal} and~\eqref{eq:bs-received-signal-pca} for No-PCA and PCA instances, respectively.
Furthermore, each sample is generated independently from each other.

The training dataset does not contain all the SNR values in the test dataset.
This is because SNR is a continuous value, with infinite values within an interval.
Additionally, SNR values may fluctuate in practical scenarios, and the detection algorithm should be able to function regardless of these fluctuations.
In other words, the DT should detect PCA even when the current SNR value is not in the training data.
Moreover, this approach uses less data to train the DT, saving computational resources during training.
This is beneficial when retraining may be needed or when large datasets need to be handled.

Therefore, the training dataset contains data with SNR varying from $ -10 $~dB to $ 30 $~dB at increments of $ 5 $~dB, while the test dataset has increments of only $ 1 $~dB.
Likewise, the number of users varies from $ 1 $ to $ M $ with increments of $ 16 $ in the training dataset, while the test dataset has increments of $ 1 $.
In both datasets, $ P_{e} $ varies from $0$ to $2.5$, with increments of $0.5$ in the training dataset and $0.1$ in the test dataset.

Finally, for each combination of $ P_{e} $, SNR, $ M $ and $ K $, were simulated $ 100 $ UL trials, with channel estimates obtained using the LS estimation technique.
When the eavesdropper power is zero, resulting in no PCA, further data is generated to balance the dataset, ensuring that an equivalent amount of data is generated for both PCA and non-PCA scenarios.

Algorithm~\ref{alg:gen_dataset} summarizes the method used to create the described datasets.
It uses the function $ sim\_uplink $ that implements~\eqref{eq:bs-received-signal} and~\eqref{eq:bs-received-signal-pca}, returning the signal received at the BS, $ \mathbf{Y} $, and the pilot symbols, $ \mathbf{x} $, employed by the compromised user.
The function $ channel\_estimation $ calculates and returns the estimated channel of the user under attack.
The $pca$ variable indicates whether there is PCA in that sample, assuming value $ 1 $ or $ 0 $ in the presence or absence of PCA, respectively.
The function $init\_blank\_dataset$ initializes a blank dataset, later populated via the $append$ function, which adds a new sample to the dataset.
The algorithm receives as input three vectors: one containing the SNR values, $ \mathbf{SNR} $, one containing the number of users connected to the BS, $ \mathbf{K} $, and another containing the transmit power used by the eavesdropper, $ \mathbf{P}_{e} $.
It also receives the number of antennas at the BS, $ M $.
Table~\ref{tab:summary-dataset} summarizes the values used to generate the samples in the datasets for a given number of antennas, $ M $, at the BS.

\begin{algorithm}
	\footnotesize
	\caption{Algorithm for training and test datasets generation.}
	\label{alg:gen_dataset}
	\KwIn{$\mathbf{SNR}$, $\mathbf{K}$, $\mathbf{P}_e$, $M$}
	\KwOut{$Dataset$}
	
	$Dataset \gets init\_blank\_dataset()$\;
	
	\For{$k \gets 0$ \KwTo $ \operatorname{length}(\mathbf{K}) $}{
		
		\For{$ i \gets 0 $ \KwTo $ \operatorname{length}(\mathbf{P}_e) $}{
			
			\For{$ l \gets 0 $ \KwTo $ \operatorname{length}(\mathbf{SNR}) $}{
				
				$ \mathbf{x}, \mathbf{Y} \gets sim\_uplink\left(M, \mathbf{K}(k), \mathbf{P}_e(i), \mathbf{SNR}(l)\right) $\;
				$ \hat{\mathbf{h}} \gets channel\_estimation\left(\mathbf{Y}, \mathbf{x}\right) $\;
				$ E \gets \lVert \hat{\mathbf{h}} \rVert^{2} $\;
				
				\lIf{$\mathbf{P}_e > 0 $}{
					$pca = 1$
				}
				\lElse{
					$pca = 0$
				}
				
				$ append(Dataset, \mathbf{K}(k), \mathbf{SNR}(l), E, pca) $
				
			}
		}
	}
	
\end{algorithm}

\begin{table}[ht]
	\caption{Summary of parameters for dataset generation for a given number of antennas $M$.}
	\label{tab:summary-dataset}
	\centering
	\footnotesize
		\begin{tabular}{ccc}
			\toprule
			& \textbf{Train} & \textbf{Test} \\
			\midrule
			\multirow{2}{2em}{\centering SNR (dB)} & 
			\multirow{2}{10em}{\centering $ \left[ -10, -5, \ \cdots, 25, 30 \right] $} &
			\multirow{2}{10em}{\centering $ \left[ -10, -9, \cdots, 29, 30 \right] $} \\
			& & \\
			\addlinespace
			\multirow{2}{2em}{\centering $ K $} &
			\multirow{2}{10em}{\centering $ \left[ 1, 16, \cdots, M \right] $} &
			\multirow{2}{10em}{\centering $ \left[ 1, 2, \cdots, M \right] $} \\
			& & \\
			\multirow{2}{2em}{\centering $ P_{e} $} &
			\multirow{2}{10em}{\centering $ \left[ 0, 0.5, 1.0, \cdots, 2.5 \right] $} &
			\multirow{2}{10em}{\centering $ \left[ 1, 0.1, 0.2, \cdots, 2.5 \right] $} \\
			& & \\
			\bottomrule
		\end{tabular}
\end{table}

\subsection{Training and Tuning the Model}

We can specify the depth of the DT to locate the best model for our problem.
To achieve this, we conducted a grid search with a stratified $10$-fold cross-validation using the training dataset to identify the best depth between one and five.
The accuracy and recall metrics were evaluated during this grid search procedure to obtain the most suitable DT.

In detecting PCA, the aim is to increase true positives while reducing false negatives, given that PCA is present.
Hence, maximizing recall is crucial for PCA detection.
Conversely, when there is no PCA in the signal, the aim is to maximize true negatives while minimizing false positives.
Therefore, it is essential to maximize precision.
Similarly, we also evaluate the F1-score, whose maximization would include maximizing both the precision and recall.

After determining the appropriate DT depth, we train the DT model using the complete training dataset.
We evaluate the resulting model under various scenarios using the test dataset.
It is imperative to note that assessing the system using the test data produced using the approach detailed in subsection~\ref{sec:dataset-methodology} is equivalent to a Monte Carlo simulation with $100$ trials.

\subsection{Baseline comparison}

We use the LRT as a baseline to compare our algorithm performance.
This algorithm consists in comparing the channel estimated energy $ \lVert \hat{\mathbf{h}}_{k} \rVert^{2}$ to a threshold $ \eta $ that is obtained by
\begin{equation}
	\label{eq:lrt_thr}
	\begin{aligned}
		\eta = \frac{M}{P_{e}} &\left( 1 + \frac{\sigma^{2}}{N} \right) \left( 1 + P_{e} + \frac{\sigma^{2}}{N} \right) \\
		&\ln\left( \frac{1 + P_{e} + \frac{\sigma^{2}}{N}}{1 + \frac{\sigma^{2}}{N}} \right),
	\end{aligned}
\end{equation}
whose derivation can be seen in Appendix~\ref{app:lrt}.

From~\eqref{eq:lrt_thr}, we can see that the optimal threshold requires the knowledge of $ \sigma^{2}$ and $ P_{e} $.
However, if the BS knows $ P_{e} $, it could use this information to directly determine the presence of PCA when $ P_{e} > 0 $.
Additionally, whereas $ \sigma^{2} $ can be estimated, estimating $ P_{e} $ can be proven difficult and unfeasible.
To establish the baseline comparison, it is reasonable to adopt $ P_{e} = 1 $, the same transmission power of the legitimate users, resulting in a method similar to the one in~\cite{Hassan2020}. 

The hypothesis testing can, thereby, be written as
\begin{equation}
	\lVert \hat{\mathbf{h}}_{k} \rVert^{2} \mathop{\gtreqless}_{\mathcal{H}_{0}}^{\mathcal{H}_{1}} \eta
\end{equation}
where $ \mathcal{H}_{0} $ is the null hypothesis, no PCA, and $ \mathcal{H}_{1} $ is the alternative hypothesis, having PCA.

\section{Results and Discussion}
\label{sec:results}

In this section, we evaluate the model in terms of detection probability.
First, we present the grid search results and find a DT for the following evaluations.
Then, we analyze the impact of four variables on the model and LRT performance: the SNR, the number of legitimate users connected to the BS and sending pilot signals, the eavesdropper transmission power and the number of antennas at the BS.
The goal is to evaluate the impact of these variables on the model's detection probability.
The SNR is defined as
\begin{equation}
	\operatorname{SNR} = \frac{P_{k}}{\sigma^{2}} = \frac{1}{\sigma^{2}},
\end{equation}
since we fixed $ P_{k} = 1 $ for all $ k $ in every scenario.

We also consider $ M = 256 $ for all scenarios, except when evaluating the impact of the number of antennas at the BS.

\subsection{Grid Search and Cross-validation}

Initially, we conducted a stratified 10-fold cross-validation grid search to locate an appropriate DT depth for PCA detection. 
The DT is grown by splitting input variables based on the Gini impurity index using the CART algorithm~\cite{breiman1984}.

The average and standard deviation of accuracy and precision for each DT are shown in Table~\ref{tab:cv_results}.
The F1-Score obtained for all depths was $0.998 \pm 0.003$, hence not included in that Table.
It is evident from the table that accuracies are better for DTs with a depth of 3 or greater, since the standard deviation is less than that of depths 1 and 2, even though the average accuracy is the same.
Regarding precision and recall, the DT depth 3 achieved the best performance, with the lowest standard deviation.
As the F1 score was the same for all DTs, the best depth would be 1.

\begin{table}[ht]
	\caption{Mean and standard deviation of accuracy, precision, recall and F1-Score obtained in the 10-fold cross-validation.}
	\label{tab:cv_results}
	\centering
	\footnotesize
		\begin{tabular}{cccc}
			\toprule
			\textbf{Depth} & \textbf{Accuracy} & \textbf{Precision} & \textbf{Recall}\\
			\midrule
			\multirow{2}{2em}{\centering 1} & \multirow{2}{6em}{\centering $0.998 \pm 0.004$} & \multirow{2}{6em}{\centering $0.997 \pm 0.008$} & \multirow{2}{6em}{\centering $0.998 \pm 0.001$} \\
			\addlinespace
			\multirow{2}{2em}{\centering 2} & \multirow{2}{6em}{\centering $0.998 \pm 0.004$} & \multirow{2}{6em}{\centering $0.997 \pm 0.008$} & \multirow{2}{6em}{\centering $0.998 \pm 0.001$} \\
			\addlinespace
			\multirow{2}{2em}{\centering 3} & \multirow{2}{6em}{\centering $0.998 \pm 0.003$} & \multirow{2}{6em}{\centering $0.997 \pm 0.005$} & \multirow{2}{6em}{\centering $0.999 \pm 0.001$} \\
			\addlinespace
			\multirow{2}{2em}{\centering 4} & \multirow{2}{6em}{\centering $0.998 \pm 0.003$} & \multirow{2}{6em}{\centering $0.997 \pm 0.005$} & \multirow{2}{6em}{\centering $0.999 \pm 0.002$} \\
			\addlinespace
			\multirow{2}{2em}{\centering 5} & \multirow{2}{6em}{\centering $0.998 \pm 0.003$} & \multirow{2}{6em}{\centering $0.997 \pm 0.005$} & \multirow{2}{6em}{\centering $0.999 \pm 0.001$} \\
			& & & \\
			\bottomrule
		\end{tabular}
\end{table}

\begin{figure}[ht]
	\centering
	\includegraphics{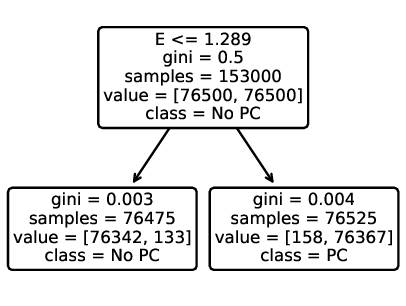}
	\caption{The resulting DT of depth one, showing the threshold of $1.289$. Larger values of energy of the channel estimate are classified as containing PCA.}
	\label{fig:decision-tree}
\end{figure}

Considering the F1-score performance and the standard deviation for the other metrics, it seems reasonable to infer that both accuracy, precision and recall do not significantly increase by increasing the DT depth.
This is so because, considering the standard deviation, there are overlaps in the results shown for each depth.
Therefore, the DT with a depth of 1 would be the best choice, since it reduces the risk of overfitting and the model's complexity without a significant drop in performance.
This choice enables quick PCA detection and countermeasure, which helps to achieve the high throughput required by 5G systems.
Moreover, the only assessed feature in this resulting DT is the channel estimate energy, $\lVert \hat{\mathbf{h}}_{k} \rVert^{2}$.
We compare this model to the LRT method presented in Section~\ref{sec:model}, which also uses the channel estimate energy.
Thus, we will use the DT with depth $1$ henceforth, which, for completeness purposes, is shown in Fig.~\ref{fig:decision-tree}, which also shows the threshold computed during the training using all training data, which is $ 1.289 $.

\subsection{Impact of the SNR}

\begin{figure}[ht]
	\centering
	\includegraphics{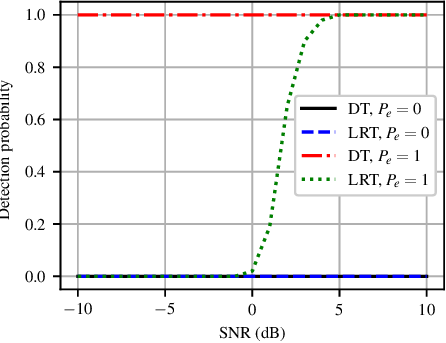}
	\caption{Probability of detection of the DT and the LRT in the presence ($P_{e}=1$) and absence ($P_{e}=0$) of PCA for different values of SNR and a 256 antennas BS.}
	\label{fig:pd_x_snr}
\end{figure}

Fig.~\ref{fig:pd_x_snr} displays the probability of detection attained using both the DT and the LRT for a BS equipped with $M = 256$ antennas and $K = 64$ connected users.
Note that the plotted data contains SNR values absent in the DT's training dataset but present in the test dataset, enabling us to evaluate the DT's performance for unseen SNR values.
Furthermore, we have plotted only the $P_{e} = 0$ and $P_{e} = 1$ values.
Notably, $P_{e}=1$ is the same transmission power used by legitimate users and used for LRT threshold calculations.

We see that the DT detects the presence of PCA ($P_{e} = 1$) for all values of SNR, whereas the LRT cannot detect the presence of PCA when the SNR is lower than $ 0 $~dB.
The probability of detection for the LRT increases for higher values of SNR, reaching $ 1 $ at SNR around $ 5 $~dB.
It is interesting to recall that the LRT requires the knowledge of the noise variance to calculate a threshold (see~\eqref{eq:lrt_thr}).
By contrast, the DT can recognize noisy cases without knowledge of the noise variance.
During training, the DT is presented with noisy data with different noise variances, not the noise variance values.
Note that both techniques have zero probability of false alarm (the probability of detection when $P_{e}=0$).

\subsection{Impact of the number of users}

We also evaluated the influence of the number of connected users $ K $ on the algorithm's performance.
In this scenario, we kept $ M $ fixed at $ 256 $ antennas and the SNR at $ 10 $~dB.
Note that the DT was trained on $K$ values ranging between $1$ and $256$ in increments of $16$.
However, we consider $K$ using increments of $4$ to evaluate the number of users not present in the training dataset.

\begin{figure}[ht]
	\centering
	\includegraphics{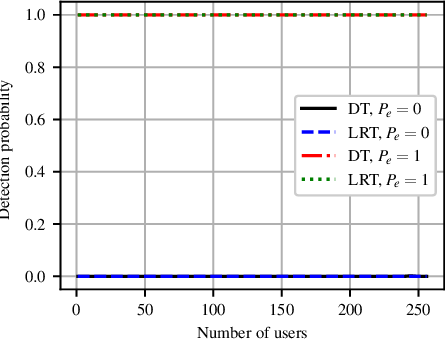}
	\caption{Probability of PCA detection for with 256 antennas BS and 10 dB SNR.}
	\label{fig:pd_x_K}
\end{figure}

Fig.~\ref{fig:pd_x_K} depicts the probability of detection with respect to the number of users for both the DT and the LRT.
We observe that both techniques could detect PCA for all the examined $ K $ values.
Additionally, both techniques presented zero probability of false alarms.

\subsection{Impact of the eavesdropper transmission power}

\begin{figure}[ht]
	\centering
	\includegraphics{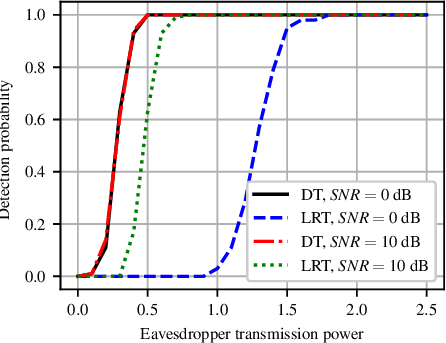}
	\caption{Probability of detection of the LRT and the DT for different values of eavesdropper transmission power.}
	\label{fig:pd_x_pe}
\end{figure}

The ability to detect PCA in both techniques with respect to the eavesdropper power is shown in Fig.~\ref{fig:pd_x_pe} for SNR at $0$ and $ 10 $~dB.
In this evaluation, we set the number of antennas as $ M = 256 $ and the number of users as $ K = 64 $.
In a noisy environment (SNR $ 0 $~dB), the DT is the first to detect PCA as the eavesdropper power increases.
The DT's detection probability reaches one for $P_{e}=0.5$.
On the other hand, the LRT probability of detection only begins to increase as $P_{e}$ approaches $1$ and reaches one for $P_{e}$ above $1.5$.

In a low-noise scenario with SNR of $ 10 $~dB, the LRT reaches a probability of detection $ 1 $ for an eavesdropper power lower than $ 1 $.
Nonetheless, the DT performance remains the same as for SNR of $ 0 $~dB, detecting the PCA even when the eavesdropper transmits at low power and regardless of the noise level.

\subsection{Impact of the number of antennas at the BS}

Next, we investigate the effect of the BS's number of antennas on the probability of detection.
For this analysis, we fix the number of users at $K = 64$ and the SNR at $10$~dB.
The DT algorithm is trained and tested individually for each value of $M$, varying from $64$ to $256$ in increments of $16$.
This is because the number of antennas of a BS is constant, hence, it is reasonable to train the DT specifically for that BS.

\begin{figure}[ht]
	\centering
	\includegraphics{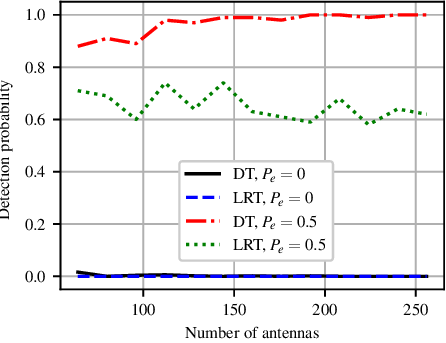}
	\caption{Impact of the number of antennas at the BS in the probability of PCA detection for an MMIMO system with 64 users and 10 dB SNR.}
	\label{fig:pd_x_M}
\end{figure}

Fig.~\ref{fig:pd_x_M} demonstrates both the DT and the LRT performance.
We notice that the probability of detection begins to decrease when the number of antennas is less than $150$ and $P_{e}=0.5$ for the DT.
In the absence of PCA, i.e., $ P_{e} = 0 $, we perceive a slight rise in the probability of detection for $M<80$, i.e., an increase in the false-positive rate.

This does not occur for the LRT, which maintains a zero false-positive rate for all values of $ M $.
However, for $P_{e}=0.5$, the probability of detection remains below $0.75$ for the most number of antennas considered.

\subsection{Comparison between LRT and DT}

It is possible to see from the results that the DT has a better detection performance in terms of detection probability.
However, both have the same probability of false alarm, which can be seen from the curves where $ P_{e} = 0$.
In this case, any positive value would mean that the PCA detector raises an alarm when the eavesdropper is not transmitting and, thus, there is no PCA.

One could use the receiver operator characteristic (ROC) curve and its area under the curve of each detector to understand the reason for this behavior.
However, the detector proposed in this work and the LRT use the same decision statistic, the channel estimate energy $ \lVert \hat{\mathbf{h}}_{k} \rVert^{2} $.
Hence, the ROC is the same for both detectors.
However, they compute different thresholds, yielding different performances after the threshold is chosen.

To understand the DT and the LRT's different performances, we recall that the LRT computes a different threshold for each value of $ \sigma^{2} $, i.e., different SNR conditions produce different threshold values.
Hence, if the SNR changes, the LRT detector computes a new threshold.
On the other hand, the DT can compute a single threshold, which is used for all SNR conditions.

\begin{figure*}[ht]
	\centering
	\includegraphics{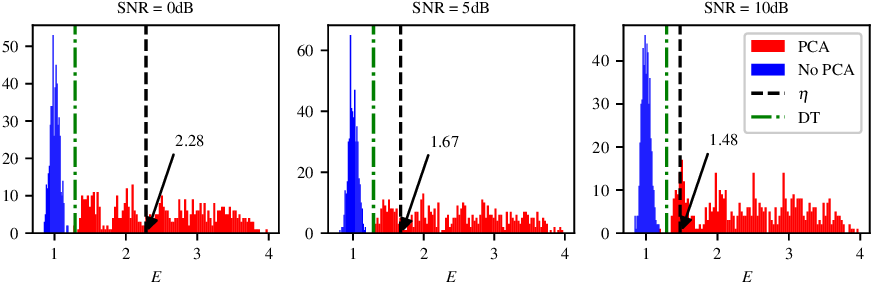}
	\caption{Distribution of data with and without PCA for different SNR environments and the thresholds calculated by the LRT (dashed line) and the DT (dash-dotted line). It is possible to see that the DT uses the same threshold for all SNR values.}
	\label{fig:hist_snr}
\end{figure*}

Considering this, Fig.~\ref{fig:hist_snr} helps understand why the LRT has the detection probability reduced for low SNR and $ P_{e} $ values.
This Figure shows the histogram of the values of $ E = \lVert \hat{\mathbf{h}}_{k} \rVert^{2} $ for different SNR values, considering some possible values of $ P_{e} $ in the scenario where $ M = 256 $ antennas.
The data shown in blue are the values of $ E $ when there is no PCA, while the data in red corresponds to the presence of PCA.
It also shows the threshold computed by the LRT using~\eqref{eq:lrt_thr} in the black dashed line.
The DT threshold is shown using the green dash-dotted line.

The first thing we can observe is that as the decision on the presence of PCA happens when the observed value of $ E $ is greater than the threshold, the LRT never raises a false alarm as the blue data is always on the left of the threshold.
However, as the SNR reduces, the detection probability reduces, since the threshold assumes higher values and the amount of data containing PCA at the left of the threshold increases.
In other words, there are more scenarios in which the LRT does not detect the presence of PCA as the SNR reduces.
It is also possible to see this happening for low values of $ P_{e} $.
Since the threshold in the LRT is calculated by assuming $ P_{e} = 1 $, scenarios where $ P_{e} $ is lower than that but greater than zero have a higher probability of deciding for the absence of PCA.
This supports the behavior seen in Fig.~\ref{fig:pd_x_pe}.

On the other hand, the DT of depth one computes a unique threshold using all the available data.
The threshold obtained in the experiment for $ M = 256 $ antennas and the training data with the parameters shown in Table~\ref{tab:summary-dataset} resulted in a threshold of $ 1.289 $.
The amount of data containing PCA at the left side of this threshold is small and, thus, the DT can provide a zero probability of false alarm and a good detection probability.
The DT fails only when $ P_{e} $ is small, in which case there is more data with PCA at the left side of the threshold.
This also supports the DT behavior as seen in Fig.~\ref{fig:pd_x_pe}.

\subsection{Computational Complexity}

To properly compare the computational complexity of the DT and the LRT, we also consider the DT of depth one, which has been the one used in this work.
Hence, once trained, all the DT does is compare the value of $ E $ to a threshold, which is the complexity of order $ \mathcal{O}(1) $.
The DT, however, requires previous training, where samples from different scenarios are presented, and the learning algorithm obtains the threshold.
The computational complexity of the training stage of the DT of depth one with only one feature is of order $ \mathcal{O}( P \log P ) $, where $ P $ is the total number of samples used to train the DT~\cite{sani2018}.

Hence, one might say that the DT requires more computational effort than the LRT.
Nonetheless, the DT can be trained offline, and the computational complexity to perform detection is, in the end, the same as the LRT.
If one considers environments in which the SNR may vary, the LRT needs to compute a new threshold for each time the SNR changes, which requires the continuous estimation of the noise variance $ \sigma^{2} $.
Thus, it is possible to affirm that the computational complexity of the LRT during the detection stage could be slightly higher than for the DT, which does not require the knowledge of $ \sigma^{2} $ for PCA detection.

\section{Conclusion and Future Works}
\label{sec:conclusion}

Multi-user MMIMO is a vulnerable communication system susceptible to malicious active attacks, including PCA that contaminate the BS's channel estimation stage by transmitting valid pilot symbols.
Due to numerous advantages, MMIMO is still the preferred technology for 5G networks.

This paper presented an ML technique using DT for PCA detection and demonstrated how to generate data to train this model.
The results revealed that the DT has a good performance concerning the probability of detection, successfully detecting PCA in both noisy and low eavesdropper transmission power scenarios.

An important aspect of the DT, and any supervised ML model, for PCA detection regards the amount of data necessary to train the ML model.
Collecting the signals to train the model in the on-scene set can be arduous and time-consuming.
Consequently, it is valuable to explore models with lower data requirements while still yielding good performance.
One example is training reinforcement learning models with artificial data that can be strengthened or substantiated after deployment.

\bibliographystyle{IEEEtran}
\bibliography{ref}

\appendices
\section{Likelihood Ratio Test for PCA detection}
\label{app:lrt}

To derive the LRT, define the null hypothesis, $ \mathcal{H}_{0} $, as the absence of PCA, and the alternative hypothesis, $ \mathcal{H}_{1} $, as the presence of PCA.
The channel estimate under $ \mathcal{H}_{0} $ is given by
\begin{equation}
	\hat{\mathbf{h}}_{k} = \mathbf{h}_{k} + \mathbf{V}\frac{\mathbf{x}_{k}^{\Her}}{\sqrt{P_{k}}\ \lVert \mathbf{x}_{k} \rVert^{2}}.
\end{equation}
As the elements in $ \mathbf{h}_{k} $ have unitary variance, and the elements in $ \mathbf{V} $ have variance $ \sigma^{2} $, the variance of $ \hat{\mathbf{h}}_{k} $ under $ \mathcal{H}_{0} $ is
\begin{equation}
	\label{eq:var0} 
	\sigma_{0}^{2} = 1  + \frac{\sigma^{2}}{N},
\end{equation}
recalling that $ N $ denotes the pilot sequence length in samples and that we consider $ P_{k} = 1 $ for all $ k $.

Similarly, under $ \mathcal{H}_{1} $, the channel estimate is given by
\begin{equation}
	\hat{\mathbf{h}}_{k} = \mathbf{h}_{k} + \sqrt{\frac{P_{e}}{P_{k}}}\ \mathbf{h}_{e} + \mathbf{V}\frac{\mathbf{x}_{k}^{\Her}}{\sqrt{P_{k}}\ \lVert \mathbf{x}_{k} \rVert^{2}},
\end{equation}
with variance
\begin{equation}
	\label{eq:var1}
	\sigma_{1}^{2} = 1 + P_{e} + \frac{\sigma^{2}}{N}.
\end{equation}

The LRT can then be obtained by
\begin{equation}
	p_{\hat{\mathbf{h}}_{k}| \mathcal{H}_{1}} (\hat{\mathbf{h}}_{k} | \mathcal{H}_{1}) \mathop{\gtreqless}_{\mathcal{H}_{0}}^{\mathcal{H}_{1}} p_{\hat{\mathbf{h}}_{k}| \mathcal{H}_{0}} (\hat{\mathbf{h}}_{k} | \mathcal{H}_{0}) 
\end{equation}
Considering that the channel has complex taps, and substituting the Gaussian probability density functions, we have
\begin{equation}
	\begin{aligned}    
		\frac{1}{\sqrt{\left(2 \pi \sigma_{1}^{2}\right)^{M}}} &\exp \left\{- \frac{E}{2 \sigma_{1}^{2}} \right\} \mathop{\gtreqless}_{\mathcal{H}_{0}}^{\mathcal{H}_{1}} \\
		&\frac{1}{\sqrt{\left(2 \pi \sigma_{0}^{2}\right)^{M}}} \exp \left\{- \frac{E}{2 \sigma_{0}^{2}} \right\},
	\end{aligned}
\end{equation}
which rearranging terms becomes
\begin{equation}
	\sqrt{\left(\frac{\sigma_{0}^{2}}{\sigma_{1}^{2}}\right)^{M}} \mathop{\gtreqless}_{\mathcal{H}_{0}}^{\mathcal{H}_{1}} \exp \left\{ \frac{E}{2 \sigma_{1}^{2}} - \frac{E}{2 \sigma_{0}^{2}} \right\}.
\end{equation}
Computing the natural logarithm on both sides
\begin{equation}
	\frac{M}{2} \ln \left( \frac{\sigma_{0}^{2}}{\sigma_{1}^{2}} \right) \mathop{\gtreqless}_{\mathcal{H}_{0}}^{\mathcal{H}_{1}} \frac{1}{2} \left[ \frac{E}{\sigma_{1}^{2}} - \frac{E}{\sigma_{0}^{2}} \right]
\end{equation}

\begin{equation}
	M \ln \left( \frac{\sigma_{0}^{2}}{\sigma_{1}^{2}} \right) \mathop{\gtreqless}_{\mathcal{H}_{0}}^{\mathcal{H}_{1}} E \left[ \frac{\sigma_{0}^{2} - \sigma_{1}^{2}}{\sigma_{0}^{2} \sigma_{1}^{2}}\right].
\end{equation}
Knowing that $ \ln \left( \sigma_{0}^{2} / \sigma_{1}^{2} \right) = - \ln \left( \sigma_{1}^{2} / \sigma_{0}^{2} \right)$ and that $\sigma_{0}^{2} - \sigma_{1}^{2} = - \left( \sigma_{1}^{2} - \sigma_{0}^{2} \right) $, we have
\begin{equation}
	-M \ln \left( \frac{\sigma_{1}^{2}}{\sigma_{0}^{2}} \right) \mathop{\gtreqless}_{\mathcal{H}_{0}}^{\mathcal{H}_{1}} - E \left[ \frac{\sigma_{1}^{2} - \sigma_{0}^{2}}{\sigma_{0}^{2} \sigma_{1}^{2}}\right].
\end{equation}
Multiplying both sides by $ -1 $ and rearranging the expression, we have
\begin{equation}
	\label{eq:thr1}
	E  \mathop{\gtreqless}_{\mathcal{H}_{0}}^{\mathcal{H}_{1}} M  \left[ \frac{\sigma_{0}^{2} \sigma_{1}^{2}}{\sigma_{1}^{2} - \sigma_{0}^{2}} \right] \ln \left( \frac{\sigma_{1}^{2}}{\sigma_{0}^{2}} \right).
\end{equation}

Finally, substituting~\eqref{eq:var0} and~\eqref{eq:var1} into~\eqref{eq:thr1}, the LRT test can be performed by
\begin{equation}
	\begin{aligned}
		E  \mathop{\gtreqless}_{\mathcal{H}_{0}}^{\mathcal{H}_{1}} M  \left[ \frac{\left(1 + \frac{\sigma^{2}}{N}\right) \left(1 + P_{e} + \frac{\sigma^{2}}{N}\right)}{1 + P_{e} + \frac{\sigma^{2}}{N} - 1 - \frac{\sigma^{2}}{N}} \right] \\
		\ln \left( \frac{1 + P_{e} + \frac{\sigma^{2}}{N}}{1 + \frac{\sigma^{2}}{N}} \right),
	\end{aligned}
\end{equation}
which can be simplified to
\begin{equation}
	\begin{aligned}
		E  \mathop{\gtreqless}_{\mathcal{H}_{0}}^{\mathcal{H}_{1}} \frac{M}{P_{e}} \left(1 + \frac{\sigma^{2}}{N}\right) \left(1 + P_{e} + \frac{\sigma^{2}}{N}\right) \\ \ln \left( \frac{1 + P_{e} + \frac{\sigma^{2}}{N}}{1 + \frac{\sigma^{2}}{N}} \right).
	\end{aligned}
\end{equation}

\end{document}